# QUANTUM REALISM AND ITS CONTRADICTIONS: A CONTRIBUTION FROM THE SOCIAL SCIENCES.

Henry Daniel Vera Ramírez[1]

https://orcid.org/my-orcid?orcid=0000-0002-3977-3073

**Abstract**

The text points out that one of the main contradictions of quantum realism, which is particularly relevant to the social sciences, is the tension between the existence of an observer-independent reality and the idea that this reality is mediated by the cognitive and interpretative processes of the subject. This contradiction arises from the central role of the observer in phenomena such as the collapse of the wave function and their influence on the construction of reality, which challenges the classical notion of an objective, non-perceptual nature. Moreover, the encounter between social realism and social interpretivism lies in the fact that, in the social sciences, it is also recognized that human interpretations and actions shape social reality, creating a scenario in which absolute objectivity becomes difficult to sustain. These tensions, therefore enrich the epistemological and ontological debate, showing that the contradictions of quantum realism transcend physics and connect deeply with the difficulties and dynamics of the social construction of reality.

[1] PhD (c) History and Philosophy of Heritage, Science and Technology, Universidad Nova de Lisboa-SENNOVA, h.ramirez@campus.fct.unl.pt

**Resumen**

Este manuscrito señala que una de las principales contradicciones del realismo cuántico, que resulta especialmente relevante para las ciencias sociales, es la tensión entre la existencia de una realidad independiente del observador y la idea de que dicha realidad está mediada por procesos cognitivos e interpretativos del sujeto. Esta contradicción surge del papel central que ocupa el observador en fenómenos como el colapso de la función de onda y su influencia en la construcción de la realidad, lo cual desafía la noción clásica de una naturaleza objetiva y ajena a la percepción. Además, el encuentro entre el realismo y el *interpretativismo* social radica en que, en las ciencias sociales, también se reconoce que las interpretaciones y las acciones humanas modelan la realidad social, creando un escenario en el que la objetividad absoluta se vuelve difícil de sostener. Estas tensiones, por tanto, enriquecen el debate epistemológico y ontológico, mostrando que las contradicciones del realismo cuántico trascienden la física y se conectan profundamente con las dificultades y dinámicas de la construcción social de la realidad.

.

**Brief review about Classic Realism**

Realism can be defined as a tendency or disposition to conform to or accept reality as it is. In this sense, realism privileges sensory and concrete reality as opposed to ideal and abstract reality. In philosophy, the name "realism" is given to a doctrine that admits and recognizes the existence of autonomous realities and, therefore, realities that are independent of the spirit that thinks and knows them (metaphysical realism) (Messer, 1923).

In this sense, knowledge is defined, in general, in opposition to phenomenalism, as a complete apprehension of what exists or is (gnoseological realism). According to one view, an *adaequatio intellectus et rei,* to the extent that intelligence, in order to know, adapts or constructs a reality that ontologically pre-exists, there is a phenomenon of interaction as a necessary condition with the existence of an observer.

Thus understood, realism gives greater rigor to the study of the world and rejects the metaphysical priority of being over knowing, both at the level of the subject who knows and of the object known. It is evident that to know, the subject must be, and it is his type of being those conditions his way of knowing. In turn, the object can be known if and only if it is, whether its reality is sensory or ideal, concrete, or abstract. This position implies an opposition between idealism, which sees in thought a principle of pure *constructive spontaneity*, and the object that turns out to be only a simple projection. For realism, on the contrary, thought is a mode of expression of a reality that sustains and informs. Idealism, by contrast, affirms that everything can be set by thought and all reality is reduced to perceiving or being perceived, while realism recognizes the existence of realities independent of the fact of being or not being known.

More precisely, realism considers the existence of realities independent of experience, both internal (consequences, states, and acts) and external (objects of the sensory world), which exist in the true sense of the word and, therefore, possess a metaphysical act of their own by which their causes and the acts by which they may eventually be known or thought are simultaneously distinguished.

Philosophical realism is not an arbitrary or gratuitous attitude, but it accepts a level of reflection by which the spirit, concentrating on itself and its acts, recognizes itself as a potential natural unity ordered to receive a reality that envelops and transcends it. In particular, the reality of the external world can be simply admitted based on the natural conviction that things are as they manifest themselves in sensory perception, which could generate naïve realism as opposed to another grounded, critical realism. The latter is based on logically prior truths (mediate realism) or on the value of sensible perception itself (immediate realism).

Historically, the term "realism" was first applied in Medieval Scholasticism to designate the view that, contrary to what occurs in nominalism and conceptualism, it gives an ontological value to universal concepts (genera and species). The exaggerated realism, or *ultrarealism,* of the Platonic perspective is attributed to universal concepts and an autonomous existence (*ante rem);* in some way, to each concept corresponds a different reality (according to Scotus Eriugena and Guilherme de Champeaux). Moderate realism, of Aristotelian inspiration, considers that universal concepts show only a common structure or essence identically realized in various individuals (*in re*). In this way, the content, which is made up of several abstract concepts or ideas (*id quad cogitatur*), corresponds to the same concrete *res* (Santo Tomás, I, q. 85 a. 2 ad-2; Noel, 1925; 1939; Van Riet, 1946; Hayen, 1954).

As a general theory of knowledge, realism manifests itself in full force against modern idealistic subjectivism. The splitting of reality between two independent subjective spheres—the subjective (*res cogitans*) and the objective (*res extensa)*— operated by the Cartesian *cogito*, had found in the problem of the bridge the problem of the passage from consciousness, or the subjective world of ideas, to the reality of the external world. If the spirit is endowed with innate ideas, it cannot ontologically determine whether a reality corresponds to it. For example, the realist attitude of Descartes does not seem to be solidly founded, since to pass from ideas to things, it is impossible to make use of ideas alone, and it is impossible to obtain reality from a theory of ideas (Roland-Gosselin, 1932; Vancourt, 1944; Brunner, 1943).

For this reason, Berkeley concludes, with logical rigor, that being is reduced to perceiving (subject) or to being perceived (object), which is not a simple idea. Kant's empirical or phenomenal realism considers that the transcendental idealism of a priori forms and the existence of a noumenon, or the *thing-in-itself,* makes reality mysterious and unknowable. The elimination of a residue of realism leads to absolute idealism, which is characterized by the absorption of being by knowing. Along with a strong reaction of volitional character (Picard, 1946), the conviction gradually insinuated itself into minds that, to recover a realistic perspective, it is necessary to abandon the idea of a Cartesian consciousness, hermit-like and isolated from one's own body of the world. In this new orientation, not everything was free of idealist influences, for example, the perspectives opened by Brentano and later explored by Husserl, Meinong, Scheler, Hartmann (1931), Heidegger, and Merleau-Ponty (1908-1961).

Due to these analyses, these authors converge in the classical idea of knowledge as intentionality or original openness of the spirit to being, which allows it to overcome in a more complex way the consciousness-world separation. However, the metaphysical-realist tendency has gained ground since the 19th century. In 1903, Moore published *Refutation of Idealism,* which is based on the neo-realist movement of the philosophers of matter and the philosophers of vitalism and existence. In France, the philosophy of the Spirit, without ignoring its influence and the diversity of its orientations, is based on the reality of the spirit or the inner life. On the other hand, the idea of a fierce defense of the realist principles by most of its representatives—despite Gilson's (1936; 1943) reservations toward critical realism, considered by him as *nonsense* should have more acceptance than a realism critically founded on reflective action. These generalities about the development of some trends of realist thought lead us to the related question of the relationship between realism and physics.

**Realism and Physics**

Until Galileo, mechanics dealt exclusively with the use and construction of instruments designed to overcome natural problems. It also included a theoretical reflection on these instruments and the effects they produced, but its purpose was manifestly technical and, as such, different from that of the natural sciences. The object of study was, then, impassive artificial movement, impossible to be produced through the natural capacities of man, since it depended entirely on an external force. These were the guidelines established, for example, in Aristotle's Mechanical Problems, his first treatise on mechanics, and therefore a starting point for the development of this discipline (Aristotle, Mechanical Problems, [2013], p. 11).

A fundamental break in the division between classical and quantum mechanics is manifested in the emphasis within physics on the concept of continuity and the wave-like behavior of the electron, which introduced the notion of a discontinuity in physical processes—suggested by the observation of discrete spectral lines in different experiments with light— and which has broadened the perspective of the electron as a particle. This in turn, supported an apparent stable situation in space-time in a classical sense. In the words of Holton (1985):

*the philosophical viewpoints about the nature of the physical world and our knowledge of it, deriving from quantum mechanics, are of considerable interest, the more so as this is still an area of vigorous research. For example, quantum mechanics has revived the…question: Do physical objects have an existence and properties entirely independent of human observer?* (Holton, 1985, p. 496).

For Born (1926), the application of quantum mechanics to an interpretation of its concepts in classical terms was necessary. He held the idea that the electron was a *particle* and that the wave function represented the probability that the electron was located at a certain point in space. Since the value of the wave function is a complex number, it can only be resolved as real through its square. The transformation of real values of the wave function would allow a proportionality in the identification of the electron at a position x, y, z through probability. If the complex number ψ is expressed as the sum of a real part and an imaginary part $\psi = a + \sqrt{-4}b$, then when squared, its absolute value will be given by $|\psi|^2 = a^2 + b^2$ which is a real number (Holton, 1985, p. 496). Born argued that the contrary assumption (adopted at first by Schrödinger), namely, that the electron is really a wave spread out in space with a charge density given by the *wavefunctions,* is inadmissible (Holton, 1985, p. 496). Some of the objections to the uncertainty principle are *i)* Only work with concepts that have operational definitions. Time would be wasted in the construction of theories based on hypothetical entities whose properties cannot be experimentally measured.

*ii)* A total theory of the electron has not been constructed. For Heisenberg (1927), *the path of* ***an electron*** *comes into existence only when we observe it* (Heisenberg, 1927, p. 185).

The uncertainty principle seems to align with the point of view of some philosophers close to the philosophical tradition of *subjective idealism*, according to which the real world consists of a set of perceptions of the observer, and physical objects have no existence or properties beyond observation. A comparable situation can be observed in the behavioral sciences: in every psychological experiment, the subject knows that their behavior is being observed and may not act in the same way as if they were not being observed.

Perhaps even idealized behavior is considered to exist when the psychologist is not present. Another powerful analogy has to do with intelligence tests, which are a measure used to identify individual properties that are likely to depend on cultural and biological circumstances— and whether intelligence is something related to an objective existence or rather an operational definition of a particular ability (Holton, 1985, pp. 498-499).

Quantum mechanics has stimulated much discussion by philosophers who claim the adoption of this theory —and the acceptance of some doctrines close to idealism— is to the detriment of realism. Abandoning, however, the idea of a real world "out there" and even the fact that the measuring apparatus is strictly inseparable from the things that are measured, does not make much difference in practice. One explanation, that accounts for the problem of measurement in quantum physics is the observation of the behavior of the photon and how the apparatus can measure it.

Rae (1986) begins with the analysis of a light beam split into two polarization components by a device in which a crystal calcite will plays an essential role. In this set up, with one incident and two emerging light beams, polarized rays are shown in horizontal and vertical (HV) orientation. This orientation should therefore be considered as ±45°, meaning that the emerging beams are polarized at ±45 on the horizontal axis, respectively. However, this form of polarization of the light beams cannot be easily explained without the following figure. If an unpolarized light beam is incident on a polarizer, and the two input beams are directed on detectors capable of counting individual photons (the photons must emerge in one or the other direction from the channels) (Rae, 1986, p. 20). Polarization is a property that can be attributed to photons because each photon that emerges from the first polarizer as vertically or horizontally polarized passes through the same channel of the subsequent HV polarizer (Rae, 1986, p. 21).

**Figure 1**

*Polarization of light beams*

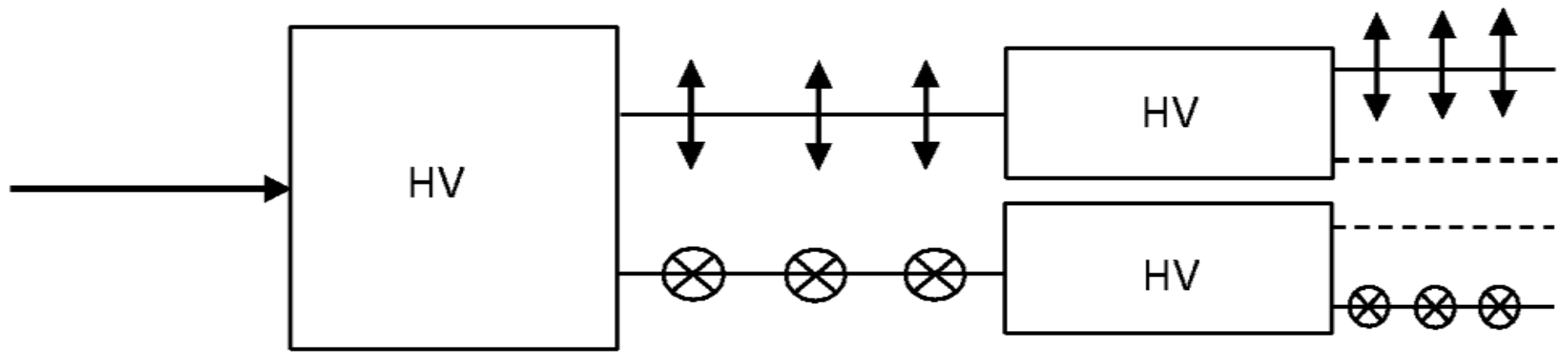

Source: Rae, 1986.

In this perspective, we contemplate the possibility that the photon passing through detectors can be emitted either vertically or horizontally. Following Rae (1986), we can imagine a set of four separate devices (including detectors) for HV polarization measurement; a photon passes through one set of devices. Each photon is in a 45-degree polarization state before measurement. The experiment is then repeated with other photons and the other three sets of apparatus, and the derivation of the idea of the universe is considered to have an impact on the sequence of operations. Initially, there is only one universe in which none of the four devices has registered a photon. This is described by the symbol OOOO located to the left of the lattice presented in the following figure:

**Figure 2**

*Multiplication of universes*

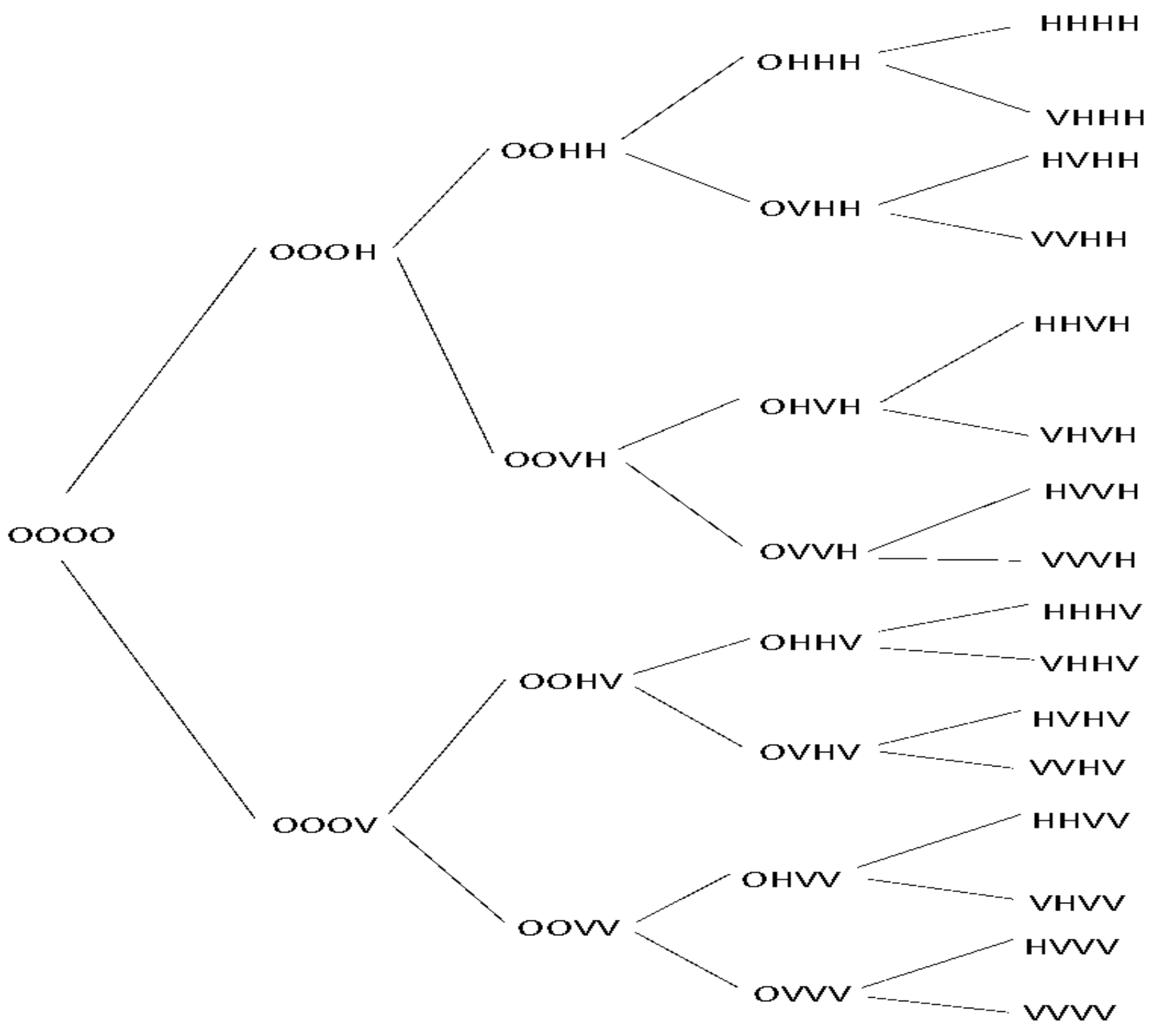

Source: Rae, 1986.

Suppose now that one of the photons passes through one of the polarizers; this has the effect of splitting the universe into two, one of which has its detectors in the OOOH state and the other in the OOOV state. If you keep passing the photons one at a time until all four have been measured, you have a total of sixteen universes represented in all permutations of H and V, as shown on the right. If one examines how many times the H and V photons were recorded as 2H and 2V, there are four records of 1H+3V, four of 3H +1V, and only one universe each was either 4H or 4V was observed. The crucial point here is that this is the same conclusion one would draw from the standard analysis of this experiment. A photon at 45 degrees passing through an HV polarizer is expected to have a 50 percent probability of being detected in either channel, but only if all four photons do not deviate from this ideal behavior (Rae, 1986, pp. 77-79).

The multiverse perspective is based precisely on the idea of a multiple reality that operates in a *parallel* way and raises a profound question and critique of the foundations of realism—not without noting that this perspective allows a significant advance in adopting a view that is not necessarily probabilistic, which would contradict the probabilistic perspective of reality proposed by positions such as Heisenberg's, for example.

Thus, as Heisenberg himself pointed out, *probability* acquires a certain intermediate reality, not unlike that of the potentiality (possibility or tendency for an event to take place) in Aristotle's philosophy. An isolated system which is evolving in this way without being observed is, according to Heisenberg, *potential but not real* (Heisenberg, 1955, p. 27. In Pauli, W *et. al.*). This contradiction is expressed in the so-called Copenhagen and Stochastic Interpretations of Quantum Mechanics, which argue that the neutron can only travel in one of the wave packets, losing its particle form and, therefore, its matter form.

Following some of Vigier's ideas (1986):

*…one deduces that between the source and the detectors, each neutron manifests itself as a wave (because of the interference) and as a particle (because of the loss of energy)*

*simultaneously. In other terms, each neutron is a wave and an article simultaneously. If one recalls that for the Copenhagen Interpretation, all measurements imply an instantaneous collapse of the wave packets, one could suggest that in CIQM (Copenhagen Interpretation of Quantum Mechanics), the apparition of the neutron in spin flippers I or II would instantaneously destroy* $\psi_I$ *or* $\psi_{II}$ *and thus destroy the observed interference. In SIQM (Stochastic Interpretation of Quantum Mechanics), on the contrary, the neutron moves in one packet (*$\psi_I$ *or* $\psi_{II}$*) only, the other (*$\psi_{II}$ *or* $\psi_I$*) being empty of the particle aspect of matter* (Vigier, 1986, p. 17).

This contradiction between particle–wave nature implies a difference of opinion about the material and the immaterial world, in a discussion about the problems of what is purely scientific, and it implies recognizing the existence of problems of a philosophical nature. One of these problems has to do with philosophers who are inclined towards the defense of the soul's survival in human nature, as in the case of Socrates and Plato, while another position, much more agnostic, considers a realistic perspective of the world that is independent of the observing subject (Popper and Eccles, 1984, p. 3).

For Popper and Eccles (1984), Kant had already shown that the difference between *the starry heavens —knowledge about the physical universe that annihilates the importance of a man — and the moral law within, our place in the Universe raises immeasurably his value as an intelligent and reportable being*[2]. This difference relates toa realistic perspective of the world.

[2] POPPER & ECCLES, 1984, p. 4.

What exactly is scientific realism? to answer this question, Psillos (1999) considers that three instances must be incorporated, which he differentiates as metaphysical, semantic, and epistemic. Each of these highlights a*non-realist* point of view about scientific theories (Psillos, 1999, p. xix):

*i)* The metaphysical instance assumes that the world is defined as a mind-independent structure composed of natural kinds.

*ii)* The semantic instance involves an evaluation that confronts scientific theories by observing them as truth-conducive through descriptions of their domains, both observable and unobservable. These descriptions are susceptible to being either false or true, although theoretical meanings are not reducible to statements about the behavior of observables, nor are they merely instrumental tools to establish connections between observables. Theoretical terms configure theories that are immersed in factual references. Therefore, if scientific theories are true, the unobservable entities they postulate must exist in the world.

*iii)* The epistemic instance is a mature and preferably successful endorsement of scientific theories as a good approximation to the truth about the world. Therefore, the entities posited by scientific theories are considered real.

These three (3) instances involve a critical ontological construction of reality that can be supported by the conceptual and experimental differences between the CIQM and SIQM perspectives.

**Scientific Realism and Ontology Fields**

Scientific realism can be regarded as the thesis that objects of scientific knowledge exist and act independently of the knowledge of their observers. More generally, realism seeks to settle the dispute about the existence of an object or thing in relation to a subject (e.g., universals, material objects, scientific laws, propositions, numbers, and probabilities). Bynum, Browne, and Porter (1983) identify three (3) major perspectives related to realism.

The first can be defined as Platonic-Aristotelian, which considers the existence of abstract or universal entities outside space and time, as in the case of Plato's theory of forms and of the material properties of things (matter and form). There are, therefore, two realisms: one *perceptual* and the other *scientific*. The first considers that things exist in space and time independently of our perception. The second—scientific realism—holds that scientific objects exist and act independently of scientists' observation and activity.

The reduction of perceptual realism to scientific realism is not extremely attractive from the ontological point of view, and the objects of scientific research are, therefore, forms or material entities. Inter-theoretical reductionism proposes that theory A can be reduced to theory B if both share the same ontology. However, theory A may have its own ontology, just as theory B does, which makes such a reduction much more complex.

On the other hand, these theories can generate, through fusion and complementarity, synthetic processes. For example, from theory A and theory B, theory C may emerge, with its own ontology. The resulting theory may share the same ontological field with either theory A or theory B. Following this logic, the question arises as to the relevance of reductionism, considering that many theories—especially in the field of quantum theories—correspond to different *ontological fields* (they see and explain different worlds). f this is the case, reduction from a theoretical point of view becomes impossible, and synthesis becomes a more plausible approach, even if it means that ontological unity could be lost. For example:

- Biology: Set of cells and structured organelles.
- Chemistry: Set of molecules.
- Physics: Set of atoms.
- Physics (nuclear and quantum mechanics): Set of particles.

It could be affirmed that they are distinct *ontological fields*. Even what would define each ontological field is related to an attribute inherent to the objects of nature: their size. Hence, the ontological field of physics, and especially that of quantum mechanics, is self-privileging because it is directed towards basic and *essential* structures. In the social sciences, these ontological fields take on other dimensions:

- History: facts (spaces, subjects, contexts, etc.).
- Sociology: actors, society, the masses.
- Anthropology: social groups, primitive societies, tribes, belief systems.
- Economics: producers, consumers (from a neoclassical approach).
- Psychology: subject, individual.

There are no strong and unrestricted ontological boundaries in the social sciences, but their difference is based on the goal of the object of study, whether it is to describe the facts (idiographic) or to establish laws (nomothetic) in the sense proposed by Wallerstein (1996).

In this context, where the existence of different ontological fields is recognized, the modern perspective of scientific realism is not reduced to a fixed position regarding perceptual and universalist influence, since the real is opposed or contrasted with the apparent or imaginary, which has led to the consideration of the study of realities that are circumscribed under apparent structures such as the material. However, this real/apparent/imaginary separation cannot be established empirically, since it presupposes a determinism in which the scientist believes, updating his theories and ensuring that they are dependent referents of reality. Thus, from the philosophical point of view, realism does not impose a perspective of interpretation of any scientific theory, which allows, for example, quantum mechanics and classical mechanics to coexist.

Realism has radically opposed nominalism, conceptualism, and resemblance theory, as well as perceptual idealism, Hume's skepticism, Kant's transcendental idealism, Mach and Avenarius's neutral monism, and the *sense-datum* analysis of logical positivism[3]. Another perspective is that of Locke (1632-1704), Galileo (1564-1642), Descartes (1596-1650), and Newton (1642-1727), who distinguished between primary and secondary qualities of bodies. There is, therefore, a difference between the physical properties that the theories suggest the bodies present and the properties that they have according to a *virtual* perception of them[4]. Modern realism has taken from its perspective a set of positions remarkably close to that of *logical positivism*, among which the perspective of the Networks occupies a key place:

For Networks, and especially in the works of Duhem (1861-1916) and Quine (1908), a holistic version of knowledge is suggested as a *force field* effect in which conditions and experiences are chained together. Building on this general idea, Hesse (1924) has argued that scientific language should be seen as a constantly growing dynamic system that enables a metaphorical extension of natural language. There is, therefore, a predicative dependence of the physical, sentient, and/or instrumental worlds and their objects on linguistic expansion, which implies an expansion of the theory-language relation and, thus, of the object-world relation. Duhem (1908), based on an anti-instrumentalist perspective, affirms:

> *The astronomer's job consists of the following: to gather together the history of the celestial movements by means of painstakingly and skillfully made observations, and then-since he cannot by any line of reasoning reach the true causes of these movements-to think up or construct whatever hypotheses he pleases such that, on their assumption, the same movements, past and future both, can be calculated by means of the principles of geometry…It is not necessary that these hypotheses be true. They need not even be likely. This one thing suffices, that the calculation to which they lead agrees with the result of observation*[5].

---

[3] In turn, instrumentalism has been opposed to realism. An example of this is the A. Osiander (1498-1522), who wrote the apologetic preface to the book by Copernicus's (1473-1453) o De *Revolutionibus* (1543), where the author of the preface considers that the heliocentric theory is a mathematical representation of the observed facts about the motion of the planets without this necessarily being true. Similarly, Berkeley (1685-1753) criticized Newtonian mechanics for its enthronement of forces as real (referring to their embodiment), which allowed a mathematical advantage over observed phenomena, facilitating their calculation. Mach and others deepened this instrumentalist perspective.

[4] Berkeley was critical of this idea, denying empirical support since all sensible perceptions of the qualities of the objects of the world would not be intrinsically independent of the mind. . Empirical realism and its ontology consider that the objects of scientific knowledge are actual and, are invariant in space and time.

What is important in this case is that there is a synchrony between calculation and observation, disregarding the feasibility of the hypotheses. Duhem says that such a physical world:

> *…at once recognizes that all his most powerful and deepest aspirations have been disappointed by the despairing results of his analysis. [For he] cannot make up his mind to see in physical theory merely a set of practical procedures and a rack filled with…[he] cannot believe that it merely classifies information accumulated by empirical science without transforming in any way the nature of these facts or without impressing on them*

[5] Quoted in Duhem (1908, p. 66), in reference to Andreas Osander.

*a character that experiment alone would not have engraved on it. If there were in physical theory only what his own criticism made him discover in it, he would stop devoting his time and efforts to a work of such a meagre importance* (Duhem,1906, p.334).

The impression left by the scientist on the accumulation of data should not simply be limited to a classification. In a certain sense, Duhem recognizes that the observer imprints a maturing character on the scientific process without falling into the extreme of subjectivity. On the other hand, Duhem (1906) affirms:

*The highest bet of our holding a classification as a natural one is to ask it to indicate in advance things which the future alone will reveal. And when the experiment is made and confirms the predictions obtained from our theory, we feel strengthened in our conviction that the relations among things* (Duhem, 1906, p. 28).

If the classification of the objects of nature is a fundamental element for the anticipation of subsequent events, experiments play a fundamental role in such classification. Later, when referring to the movement of the planet Uranus:

*If [the theorist] wishes to prove that the principle he has adopted is truly a principle of natural classification of celestial motions, he must show that the observed perturbations are in agreement with those which had been calculated in advance; he has to show how, from the course of Uranus he can deduce the existence and position of a new planet, and find Neptune in an assigned direction at the end of his telescope* (Duhem, 1906, p. 195).

Disturbances, therefore, are not synonymous with a lack of determinism, but rather signals of change in a perspective of reality that must be detected by the scientist.

**Realism and Instrumentalism**

The relationships between the observed data must generate, as a result, a logical relationship between the objects of nature. The data functions as an instrument to establish this relationship and allows us to arrive at the objects of reality through experiments. For Duhem (1906):

> *But the more complete [the theory] becomes, logical order in which the theory orders experimental law is the reflection of an ontological order, the more we suspect that the relations it establishes among the data of observation correspond to real relations among things, and the more we feel that the theory tends to be a natural classification* (Duhem, 1906, p. 26-27).

In this respect, Harré (b. 1927) considers that models have lengthened the discussion on how, in science, theoretical discussions regarding entities and processes of the scientific endeavor are initially imagined, which subsequently allows the exploration of what is observed as a phenomenon and its establishment as real through an extension of the sensible to the construction of instruments that allow the detection of the effects of the *theoretical phenomenon*. The criterion of attribution of realism is based on a causal invocation of a previously imagined model.

The historical perspective of scientific change, which includes the works of Popper, Bachelard, Koyré, Kuhn, and Hanson, analyzes the consistency and variation of the meaning of the scientific process. In the case of scientific change, Kuhn and Feyerabend have concentrated on the analysis of rival theories and scientific revolutions. Scientific objects are the result of *abrupt* changes in the forms of interpretation of scientific communities.

Bhaskar's (1944) *experimentalism* has considered that a condition of the intelligibility of experiments and the applied activity of science is that its object exists and acts counterfactually, independently of the activity. This is the case with the empirical realism of Kant, Hume, Popper, and other so-called *super idealists*. Science is an activity where experiment, and therefore its design, plays a fundamental role, and empirical realism involves an anthropocentric ontology, relegating the importance of experiment.

Empiricism based on experience is directed towards the understanding of *natural order*, while idealism would consider that this natural order is constructed by human beings. Realism is given by an assumption of causal investigation of nature that is, however, socially constructed and adapted to our cognitive possibilities.

**Putnam and Quine**

Putnam's (1926) perspective of scientific growth considers that the cumulative character of scientific knowledge is an indicator of those theories that describe natural structures, which implies a better approach to the explanation of their phenomena. This perspective is opposed to the Popperian essentialist perspective that considers reality to be the set of the most updated explanations of a field. What is interesting is that these latter explanations can be contrasted from an ontological point of view, recognizing change and growth[6].

In *Materialism and Relativism*, Putnam (1992) considers:

> *While individual philosophers continue to produce and defend as wide a range of metaphysical views as they ever have, today two outlooks have become dominant in American and French philosophy; these are, respectively, the outlook of materialism and relativism. Although few American philosophers actually call themselves materialists, and I do not know of any French philosophers who actually call themselves relativists, the terms "physicalism" and "naturalism" have become virtually synonymous with materialism in analytic philosophy, while the thrust of deconstructionist views is often clearly relativist, if not downright nihilist. I have argued for some years that both styles of thought are too simplistic to be of much help in philosophical reflection* (Putnam, 1992, p. 60).

On the one hand, the fact of recognizing that some philosophers of the French current[7]— the fact that the relativists— have dismissed the separation between physicalism and naturalism and have associated these words with materialism and analytic philosophy is a critique that would be aligned with the perspective proposed by Bouveresse (1999) in relation to the critique of these currents. But the field of the discussion on metaphysics is not reduced to this critique from analytic philosophy as a great field of action; rather, it supposes a set of assumptions supported by Putnam (1992). By the way:

> *Contemporary analysis of counterfactuals suggests that two things determine it: 1) which possible situations (or which possible worlds) are close to the actual world (or, as I would prefer to say, relevant to the statement when we consider the actual situation in which it was made) and 2) what would happen in those possible situations. For a physicalist, the latter is no problem: if the possible situations are completely described in the language of physics, say by a "state function" in the sense of quantum mechanics, what will happen in that situation (or the probability that any given thing will happen in that situation) is determined by the laws of fundamental physics (or Fodor might say that it is determined by the laws of fundamental physics plus the laws of the relevant "special sciences"). But on the notion of something being a law of physics (and/or a law of the special sciences)-not a law of the accepted physics (or, respectively, special sciences) but a law of the true physics (special sciences), whatever that may be-and this is hardly interpreted relativistically in this formulation, the relativist has a problem: the truth value of the statement that my kitchen needs painting depends (for the relativist) on the truth value of the statement that* *people (in various hypothetical situations) would say that the paint in my kitchen is dingy and peeling, and that in turn depends on what the relevant laws are (physical, biological, psychological, and so on), and that in turn depends on what people would say the relevant laws are* (Putnam, 1992, p. 70).

[6] In this respect, Nagel (1901) and his continuous perspective and Popper (1984) and his *discontinuous* perspective, point to a historical analysis of the stratification of the sciences.

[7] One of the philosophers Putnam criticizes for the relativist stance to which Putnam points, is Derrida. In this sense: "*Derrida, I repeat, is not an extremist. His own political pronouncements are, in my view, generally admirable. But the philosophical irresponsibility of one decade can become the real-world political tragedy of a few decades later. And deconstruction without reconstruction is irresponsibility*" (Putnam, 1992, p. 133).

The distinction between relativism, which recognizes that the laws of physics operate even in situations of behavior tied to probability, suggests the existence of two types of laws: the fundamental laws of physics and the laws relevant from the point of view of the specific field or of the sciences called here “special”, the crucial point of Putnam's approach. In this sense, the role of probability would come to occupy a significant place in these laws of special sciences since they would function in an analogous way to the claims established by the laws of universal character. It would seem at first glance that this division into *laws of sciences* in the form of a kind of matryoshka—contained one within the other and successively in the division by fields—would allow a better understanding of phenomena from their context. But it is worth noting criticisms, for example, from those who consider that it would be contradictory that the pretensions of a standardized model of the laws be contained or contradict themselves.

Putnam (1992) further states:

*When a French philosopher wants to know if the concept of truth, or the concept of a sign, or the concept of referring, is consistent or not, he proceeds by looking at Aristotle, Plato,*

*Nietzsche, and Heidegger, and not by looking at how the words "true," "sign,"and "refer" are used* (Putnam, 1992, pp. 120-122).

Continuing with his critique of metaphysics, he considers that French philosophers concentrate on the observations or theoretical meditations that *idealist* or *metaphysical* philosophers make on different terms or concepts, ignoring their use. What does Putnam mean by the idea of use? He refers to the word in its ordinary and contextual use beyond its *philosophical* reference. The concept of truth, or reference to which Putnam refers, can be understood as a purely abstract concept —as ideal entities— and, in this sense, outside the ordinary sense of its use or, in a much more *analytical* case, in terms of the logic under which they are integrated into reality. In this way, a reference to all objects as a way of arguing their aggregation in the world is insufficient and mistaken, without ignoring that objects are aggregations of other elements. Putnam affirms:

> *The idea that "object" has some sense which is independent of how we are counting objects and what we are counting as an "object" in a given situation is an illusion. I do not mean by this that there "really" are "aggregates," and there really are atoms and there really are sets and there really are numbers, and so on, and it is just that sometimes "object" does not refer to "all objects". I mean that the metaphysical notion of "all objects" has no sense* (Putnam, 1992, p. 133).

Referring specifically to quantum mechanics—and much more specifically to the concepts of superposition and the variables of *momentum* and position as a sum of

vectors—these involve a difficult interpretation from the point of view of realism, precisely because of their elevated levels of abstraction in relation, especially regarding the action of the observer, which is considered to play a puzzling role in its function. It is important to note how schemes such as Feynman's diagrams contribute to the clarification of the action of many particles. From this, Putnam states:

> *Again, in quantum mechanics, any two states of a system can be in a "superposition"; that is to say, any particular state of a system, involving having a particular number of particles or a particular energy or a particular momentum, can be represented by a kind of "vector" in an abstract space, and the superposition of any two such states can be represented by forming a vector sum. These vector sums are sometimes classically very difficult to interpret: what do we make of a state in which the answer to the question "How many electrons are there in the box?" is "Well, there is a superposition of there being three electrons in the box and there being seventeen"? But we can represent such unthinkable states mathematically, and we know how to derive predictions and formulate explanations using them. This principle of superposition applies to field theory as well as to particle theory: the "field states" of the quantum field theorist; they are typically superpositions of the field states of the classical theory. We may say, then (and here I leave out entirely the puzzling role of the observer in quantum mechanics), that from the point of view of quantum mechanics, the world consists of fields in "funny" states. But-and this was the discovery of Richard Feynman, it is also possible to think in a very different way. We can* *think of the world as consisting of particles (although we have to vastly increase the number of particles, we postulate in order to carry this through), and we can think of any situation that we describe in field physics as a superposition of an infinite number of different ways of thinking in quantum field theory. In one way of thinking, the way the physicist thinks when performing the usual field calculations, the system is in a superposition of field states. In the other way of thinking, the way of thinking when drawing "Feynman diagrams," the system is in a superposition of particle states. In short, the system may be thought of as consisting of either classical fields or classical particles* (Putnam, 1992, pp. 120-122).

The recognition of Feynman diagrams as analogies for the understanding of the behavior of particles introduces the perspective that physical systems, can legitimately be understood as bifurcated into two worlds or two distinct systems. On this point, Putnam states:

*Consider a given physical system which the physicist represents twice over, once in the language of fields and once in the language of particles (say, by drawing Feynman diagrams). What I am saying is that this is a real system, and that these are two legitimate ways of talking about that real system. The fact that the real system allows itself to be talked about in these two very different ways does not mean either that there is no real physical system being talked about, or that there are two different physical systems in two different Goodmanian worlds being talked about* (Putnam, 1992, p. 120-122).

On the other hand, Quine (1982), when dealing with the analysis of terms, recognize that in addition to the existence of singular and general terms, there is another classification in which concrete and abstract terms are included. Concrete terms are those that refer to individuals, *physical objects, and events,* while abstract terms refer, for example, to numbers, classes, and attributes (Quine, 1982, p. 259). Quine's proposed examples point to proper names, such as Cerberus, Earth, and the author of Waverley, as concrete, while seven (7), three (3), and four (4), piety, etc., are abstract. Again, general terms such as "man," "house," or "red house" are concrete if "house" and "man" are considered individually concrete, while others, such as "zoological species" and "virtue", are abstract. However, caution must be exercised in the way abstract terms are used, bearing in mind their general character and that their distinction from concrete objects has to do with the type of object to which they refer.

This differentiation of terms also makes it possible to point out some aspects related to the uncertainty principle and the wave function of quantum mechanics. The uncertainty principle, as a physical reality, entails a level of abstraction that delimits the capacity of observation of the phenomenon and requires recognition of the existence of hidden variables that do not allow a deep analysis of the phenomenon. On the other hand, by recognizing the duality of *momentum* and position as antagonistic measures, physical reality operates within that contradiction. Assuming an absolute measurement of a physical phenomenon is, therefore, an abstract-realistic perspective in the case of the uncertainty principle. If it is possible to measure one variable from the realist point of view, the inability or impossibility of accessing an exact and simultaneous measurement of the other variable cannot be ignored, which implies a level of realism, but not an absolute realism.

The wave function and its collapse, in which the observer seems to create nature according to the Copenhagen interpretation, would have, in this perspective, a current that would go in direct contradiction with the proposals of a reality existing independently of the observer and in whose construction the observer would not mediate, but rather act as an articulator of reality with cognitive and interpretative elements that would not necessarily reconfigure reality itself. It is at this point that Quine considers language as a powerful element for the construction of this reality.

On the other hand, Quine (1982) insists on the differentiation between singular and general, from the point of view of language, a differentiation that is vital from the logical point of view, but from the physical point of view, it would be tied to what he calls concrete terms, whose value, in the explanation of the physical universe, would have a greater significance (Quine, 1982, p. 260). In addition to this, whatever word is used to designate an object and only an object is a matter of language and is not necessarily contingent on the facts of its existence.

The use of the term "existence" versus "must be" must be resolved from a concreteness in space-time. When the Parthenon and the number seven (7) are said, they are not said to be; the Parthenon is an object that exists in a place and space-time, while the number seven has another connotation that has nothing to do with its situation of being. In contrast to seven (7) and the Parthenon, there is no such thing as Cerberus, and there is no such thing as 0/0. This difference lies in the limitation of its spatio-temporal existence. The meaning of the word Cerberus, which names an object that should be an object in spacetime, is separated from the analysis of the word Cerberus, so there is something that is called Parthenon and Bucephalus that also exists in spacetime and is called Parthenon (Quine, 1982, p. 263). The effort to preserve meaning for Cerberus through a shadowy entity that is indirectly named allows Cerberus to retain its meaning even without naming it. Words such as *"and" and "sake"* retain their strength of meaning, even without references to named objects. Even when the word is the name of something, its meaning may not appear to be identifiable with the name of the thing. An example, taken by Quine quoting Schrödinger relates to Mount Everest, which has been known as Everest and Chomolungma (in the native language), implying a synonymous meaning (Quine, 1982, p. 264).

Thus, from Quine's point of view, spatio-temporal limitations play a crucial role in the development of what are called concrete terms, which not only correspond to a linguistic expression but must be translated into physical reality under the independent existence of the one who names them. In this sense, the one who names—or the observer—directly influences by naming a physical phenomenon, but this is translated into a concreteness inherent to the physical world. If the observed and named phenomena are considered from the physical point of view, we will have strong criticism from the followers of the Copenhagen school of a version of reality that is only possible in its construction through language.

## Analogies and the Copenhagen School

The analogical approach to scientific theories can be analyzed historically from the works of Achinstein (1991) and Hesse (1953; 1966), which focus on physical models. The first problem scenario in model building is that scientists must or wish to investigate a set of phenomena or the behavior of a target physical system (x) (e.g., light waves). To this end, a theoretical model of (x) is constructed in which a set of mathematically complex assumptions are employed, which provide a starting point for the study of the behavior of (x).

The selection of modeling assumptions for (x) is guided by substantial similarities between the target system (x) and other physical systems (y). The expression that summarizes this process is *model M of (target system (x), based on (source system) (y).* But (y) is a different physical system, although with some similarities with the physical system (x). Following Hesse (1966): the relation between the source system (x) and the target system (x), is characterized by the existence of:

*a.* Some positive analogy: that is properties or relationships between properties that (x) and (y) share.

*b.* Negative analogies, i.e., some properties or relations, between properties they do not share.

c. *Some neutral analogies,* for example, some properties about which it is not possible to know the positive or negative effects of the action done on the analogy (Psillos, 1999, pp. 140-141).

The kinds of analogies that meet these criteria also account for their real capacity to explain the real world, and they imply a limit. The enormous change in the natural sciences during the last hundred years following the year 1808 may be correlated with an enormous field of advances in the sciences from different fields such as biology, chemistry, and, of course, physics. The discovery that there was a discrete quantization structure in nature was a milestone in many aspects and in the quantization of energy in systems that exhibited vibratory motion or periodic fluctuation, as in the case of the oscillator in the Planck electromagnetic field, which, as already observed, is the last bastion of these remarkable developments.

Although Dalton had shown in the 1800s that chemical matter is not an infinite and indivisible continuum and that it consists rather of atoms that maintain their integrity in chemical changes, this set of ideas would have implied, for example, the consideration of the cell theory of Schleiden (1804-1881) and Schwann (1810-1881), who revealed that life or living matter depended directly on the cell and its corpuscles. The work of Gregor Mendel (1866) followed this important trend and asserted that matter was governed by heredity that was associated with a structure of defined particles (genes), which could be transmitted from generation to generation. Meanwhile, the physics of heat, electricity, magnetism, and light had, in the 18th century, been explained by an extensive visualization of fluids (Holton, 1985, p. 444).

Sensible heat was identified with the discrete motion of atoms. The electric charge was found through to the discovery of the electron, whose motion should serve to separate magnetic fields. Finally, the energy of light and atomic oscillators was also quantized. This introduced a new point of view that pushed new fields of science and similarly moved towards new mental models of change for certain quantities and processes, a change that affected the idea of a continuum in favor of a well-defined idea of particle or quantum.

In a sense, this could be explained to mean that changes were noted in the development of better instrumentation and techniques more open to new levels of observation, strengthening the measurement of phenomena, but without resolving the intrinsic issues or the contradictions within quantum mechanics. However, this is extensively explained after the fact by the technical developments that were often stimulated by a new conceptual movement that affected the dynamics of the nature of modern science and had to do with those who followed a realist-determinist model and those who followed the Copenhagen perspective (Jammer, 1966, p. 445).

As Jammer (1966) states, although the ψ function seems to play a crucial role in quantum mechanics through the Schrödinger wave function, some physicists consider that this idea is still innocuous and superfluous, based mainly on the advances made by the theory of matrix mechanics,, proposed by Heisenberg, which contributed in that it was considered as an alternative to the set of roles played by the calculation of frequencies and intensities of spectral lines, using only relations between observable quantities. Similarly, Pascual Jordan (1902-1980) and Max Born (1882-1970), like Heisenberg, developed a mathematical method to complete the processes inside the atom, which had the same results as those obtained by Schrödinger. On both theories, Jammer stated:

> *It is hard to find in the history of physics two theories designed to cover the same range of experience which differ more radically than these two—*Schrödinger*´s wave mechanics and Heisenberg´s matrix mechanics* (Jammer, 1966, p. 495).

In the Copenhagen interpretation, the principle of complementarity may appear at first sight as somewhat vague. In addition, for some physicists, it has meant an enormous weight in the interpretation since Bohr made his presentation in 1927. His ideas, at the time, were not clearly expressed and were perceived as an initial confusion, but with time, a degree of acceptance was achieved. Copenhagen's interpretation has had many detractors who have found fault with its explanation of the microworld. However, it has remained in the domain of quantum mechanical interpretation. In agreement with this interpretation, many physicists have emphasized that such an interpretation is philosophically offensive because beyond confronting the problem of duality, Bohr's principle seems to focus more on the limitations of quantum mechanics in observing attenuated phenomena (Herbert, 1985, p. 524). In this regard, Newman (1955) has said:

> *In this century the professional philosophers have let physicists get away with murder. It is a safe bet that no other group of scientists could have passed off and gained acceptance for such an extraordinary principle as complimentary* (Newman, 1955, p. 42).

The interpretation of Copenhagen was also extended thanks to the state function of Max Born (1882-1970) and supported by Born's idea that it is related to the premise that the wave function is related to the position, probability, and amplitude of the wave. At the same time, it is easy to express this notion mathematically—but to fully understand it, one must recognize the importance of the profound significance of measurement in quantum mechanics (Newman, 1955, p. 69).

In the words of Born (1984):

> *Before we leave the Correspondence Principle, I want to add a note on **quantum epistemology.** Discussions of this principle seduce some readers into the impression that quantum mechanics somehow subsumes classical physics, that it is more fundamental and obviates the need for classical mechanics in physical theory. This is wrong. Although some* *physicists have tried to conjure up an interpretation of quantum mechanics that does describe everything* (Born, 1984).

The prevailing view that description is not desirable is that no non-classical quantum mechanics can describe the universe. It cannot subsume one into the other. The reason for this perspective that makes the dualistic view necessary lies between the lines, since the correspondence principle, which has no clear space of discussion broad enough to categorically state that it broadens Copenhagen's perspective about the description of a microscopic system, which, moreover, is not always going to be smoother within a simpler actuality, which is a feature of the classical description (Woo, 1986, p. 923, and Morrison, 1990, p. 923).

On the other hand, classical statistical mechanics attempts to derive the macroscopic properties of matter, starting with the mechanical laws and the behavior of individual particles. To describe states in equilibrium, it only requires observable measurements, which are measurable through correlations between states that are systems consisting of a substantial number of particles subject to measurement. The observables are described by continuous functions over phase space, and the states are represented by linear assignments of a number for each observable.

The derivation of blackbody radiation laws is analyzed and presented chronologically in order of Kirchhoff's theorem, Stefan's law, Wien's displacement law, and the Rayleigh-Jeans formula. The wave packet is a path representation of the wave function ψ (x, t) for a given physical system. The interpretation of the wave function ψ is given by the Born postulate:

> *The probability of finding the particle described by a wave function ψ in a small (differential) region surrounding the position x is proportional* $|\Psi|^2 \cong \Psi * \Psi$, *the square of the magnitude of the wave function, evaluated at the position x* (Thomas, 1981, p. 69).

This postulate concerning particles is analogous to the situation in electromagnetic wave propagation, where the probability of finding a photon at a point in space is proportional to the square of the intensity of the electromagnetic wave at this point. Moreover, in classical electromagnetic theory, the square of the intensity of the wave is interpreted as the measure of the energy density at the point in question.

Similarly, the time dependence of the *wave packet* seems to indicate, without doubt, a theoretical tool that serves to describe the location of the particle, which has merit in that it finally introduces some degree of localization in the infinitely extended plane of wave solutions to the Schrödinger equation. The broadening of the packet with time deservedly reflects that there is some initial uncertainty at the time of the corresponding particle expanding by *k* values, which, as time progresses, results in greater uncertainty in the particle's position. The particle itself should not be considered dispersed with time: it is only our knowledge of its location, which decreases with time.

This is a frustration from the epistemological point of view of quantum mechanics since our knowledge is restricted to what is contained in the statistical function ψ (r, t).

On the other hand, the intellectual satisfaction of quantum mechanics at this point derives from the statistical predictions of the function $\psi$ (x, t), which agree with experimental results (Thomas, 1981, p. 70).

The procedure of explaining atomic phenomena in classical terms by the superposition of additional arbitrary relations was carried further than has been indicated above, but there is no purpose in giving more details. It is clearly and intrinsically unsatisfactory to have both radiation and matter being treated sometimes as waves and sometimes as particles in an arbitrary manner, and discrete hydrogen levels being produced by *ad hoc* rules, which are completely contrary to the spirit of classical mechanics, to which they are applied. What is required is a basic reformulation of the theory in such a way that both those classical concepts which remain correct and the Planck-Bohr-de Broglie rules shall appear as natural consequences of some coherent whole. For this purpose, the above rules are an important guide to the non-classical feature.

The difficulties in understanding the experimental results that have remained since the beginning of the last century were overcome by the explanation of a narrow scheme of the atomic model and the discovery of X-rays and radioactivity. However, many of the difficulties associated with the physical phenomena to be studied were also related to phenomena, now described as the distribution of the thermal spectrum of radiation from the black body, and the low and specific temperature of solids.

A second beginning for the understanding of a second class of phenomena was the already described explanation by Planck in 1900, when he clearly stated that to explain the black body spectrum, the acceptance of this emission and absorption of electromagnetic radiation in *discrete quanta* was analyzed individually from an amount of energy E which is equal to the frequency of the radiation multiplied by Planck's universal constant *h*, $E=hv$.

In this sense, the dual character of electromagnetic radiation behaves in some situations as a wave and in others as a beam of quantum corpuscles. The existence of discrete values for the measurement of parameters of atomic systems—not only electromagnetic radiation— is confirmed through the theories of Einstein and Debye about the specific heat of solids, the classification of the spectral lines of Ritz, the experiment of Franck and Hertz on the discrete energy lost in the collision of electrons with atoms and the experiment of Stern and Gerlach, who showed that the components of the magnetic moment of an atom along a magnetic field have discrete values (Schiff, 1968, p. 2-3). The following table shows the different experimental advances that account for theoretical explanations related to discrete-value measurements and their antecedents:

**Table 1**

*Experimental advances in the measurement of discrete variables*

| Theory | Measurement |
|---|---|
| Blackbody radiation (Planck, 1900) | Electromagnetic waves |
| Diffraction (Young 1804, Love, 1911) | Electromagnetic quanta |
| Photoelectric effect (Einstein, 1904) | Electromagnetic quanta |
| Compton effect (1923) | Electromagnetic quanta |
| Combination principle (Ritz-Rydberg, 1890, 1908) | Electromagnetic quanta, discrete values for physical quantities |
| Specific heats (Einstein 1907 Debye, 1912) | Electromagnetic quanta, discrete values for physical quantities |
| Franck-Hertz experiment (1913) | Electromagnetic quanta, discrete values for physical quantities |
| Stern-Gerlach experiment (1922) | Electromagnetic quanta, discrete values for physical quantities |

Source: Schiff, 1968, p. 4.

This general perspective of the advances in experimental measurements shows considerable progress in the search for the precise measurement of the phenomena that are observed and that give context to our interpretation of reality. Although measurement scales account precisely for another form of human construction and are referenced in the analysis of these phenomena as an element of comparison and guidance, it should not be ignored that the constant character that many of these measurements have can allow us to approach reality without ignoring that the observer has a direct influence on its construction.

**Kuhn's Point of View**

For Kuhn (2001), reality is circumscribed by the idea of *natural families*, which should not be considered as an essentialist notion since the idea of reality for Kuhn underlies a coherent proposal for the interpretation of reality and its dynamics that affirms a *principle of independence* of scientific contents. The fact that Kuhn confers so much importance to what he calls internal processes as a form of interpretation of the natural world challenges his apparent approach to the principle of independence, departing from the logical-empiricist thesis (Veiga, 2008, p. 75).

Kuhn restricts the conditions of effectiveness of reality through the concept of a reality niche and, in this way, manages to reevaluate and extend an alternative to the concept of truth from an essentialist position, highlighting the *internal* as the perspective that defends that scientific theories are created based on experience and break with the idea of the establishment of an equivalence between theory and reality, which is context-independent. However, in principle, doubt arises as to what Kuhn considers internal, whether this is based on a cognitive perspective or the understanding from the perspective of a subject who constructs reality, or of a mediator of the world that manages to unite the Platonic world and the mathematical world of Penrose's perspective.

Bird's (2000) criticisms of Kuhn's ideas focus on the error of confusing the types of cognitive domains: language vs. perception and the sensory conversion of natural stimuli with an interpretation and organization of those stimuli. According to Bird (2000), there is an imprecise notion of perception since Kuhn applies two different ideas to it when referring to the imagined reality and the essential reality whose extrapolation is diffuse, as it is the objective and eco-referenced reality, which Kuhn recognizes when he states that scientific theories are more a product of the scientists' *imagination* (Veiga, 2008, p. 76).

Kuhn (2001) considers that nature *instructs* cognitive processes. For example, it only makes sense to speak of empirical anomalies [8], if there are regularities in nature. Otherwise, a high instability of natural phenomena would not allow the articulation of any structure of knowledge, neither individual nor collective. The very phenomenon of scientific legitimization of facts, problems, methods, and terms of language presupposes the existence of regularities (Veiga, 2008, p. 81).

Bird (2000), for his part, considers that there is a problem in *perception* seen from the perspective of a process of *extraction* of information from the natural world if it is understood as a process of fixed and universal rules. This process must consider the difference between seeing and being aware of the relationship between the sense organs and memorization (psychological processes), distinguish the regularity of the *rules of conversion* of information, and conclude whether one speaks of the reality experienced by others or not (Veiga, 2008, p. 82).

Kuhn resorts to the notion of *Niche* to find a balanced and sustained solution that articulates human reality, disciplinary reality, and individual realities. If the scientist has already chosen pertinent information even before a process of memory that is inconclusive for obvious reasons, what is evident is a univocal arbitrariness that is also real concerning that which the scientist considers to converges in the phenomena he analyzes: each disciplinary niche contains its reality, which comes from the conjugation as much as from the intersection and logical meeting of its inhabitants (the scientists) and of all the numbers (Platonic world) that inhabit what could be called a referential spectrum of verisimilitude which implies an essential reality (Veiga, 2008, p. 89).

[8] They could be related to the disturbances referred to by Duhem.

This essential realism, therefore, is not exempt from deep criticisms, which have arisen due to the need for a greater understanding of how data or information from the world is translated into a reality that precedes the observer.

## Decoherence and Entanglement: Haroche-Penrose's Explanation

One of the concepts that allow us to approach the theoretical body of quantum mechanics is that of the so-called decoherence. From Haroche's perspective, the concept of decoherence plays a fundamental role. The term refers primarily to an attempt to explain how an entangled quantum state can lead to a classical or non-intertwined state. It refers to how a physical system, under certain conditions, can cease to exhibit quantum effects.

This phenomenon is related to the superposition principle. If we consider the superposition principle—which is observable in systems consisting of at least two particles that interact with each other and then separate, —we can, in principle, examine the collision of two identical atoms (Haroche, 2002, p. 90).

Each of these atoms has two energy levels *e,* and atom 2 is in its ground state (*g*-state). Two things can happen during the collision: either the atoms keep their initial energies, or they exchange them. Classically, the atoms would have to *choose* one of these two possibilities. The quantum rules are different. They can follow both paths at the same time. After the collision, the system is in a superposition of the state in which atom 1 is at *e* and atom 2 is at g and the state in which atom 1 is at *g* and atom 2 is at *e*. Each of these states has a complex amplitude associated with it. The squared moduli of these amplitudes represent the probabilities of finding one or the other of these two situations during the measurement of the two atoms. Note that although the result of the measurement on each atom is random, the correlations between the results of the measurements are certain (Haroche, 2002, p. 90).

If atom 1 is at *e*, atom 2 is at *g*, and vice versa. This perfect correlation, which can be observed regardless of the type of measurement performed on the atoms, is called *entanglement*. This entanglement holds even if the two atoms have moved away from each other and are separated after the collision by an arbitrarily large distance. It describes a fundamental *non-locality* of quantum physics[9]. A measurement of atom 1 can has an immediate effect at a great distance on the result of the measurement of atom 2. There is thus an instantaneous, immaterial quantum link between the two particles (Haroche, 2002, p. 91).

[9] It was Einstein, with his collaborators Podolsky and Rosen, who first discussed this disturbing aspect of quantum theory in 1935. Since then, it has been known as the EPR paradox. For Einstein, it was a serious flaw in the theory, since it predicted effects that seemed patently absurd. The problem has since been taken up by other physicists, notably John Bell (1928-1990) in the 1950s, through experiments on entangled photons, which have shown that nature behaves

If entanglement seems strange to us, it is largely because, like quantum interference, it is never observed in macroscopic objects. This brings us to the famous metaphor of Schrödinger's cat. Reflecting on the EPR problem, Schrödinger's went even further. He wondered what would prevent a microscopic entanglement phenomenon from being amplified to involve a macroscopic system. Consider an excited atom that emits a photon when it de-excites. Quantum mechanics tells us that, before the photon is finally emitted, the system is in a superposition of a state in which the atom is still excited and one in which it has already de-excited. Each of these terms is assigned its complex amplitude in the expression of the overall state of the system. But Schrödinger's points out, a single photon can trigger a macroscopic event.

Imagine, for example, that we are locked in a box with a photon. Suppose that the photon emitted by the atom activates a device that kills the cat. If the atom is in a superposition of a state in which it has emitted a photon and one in which it has not yet emitted a photon, what is the state of the cat at that moment? If we accept that the cat can be described by a *well-defined* quantum state (and here, as we will show later, we touch as prescribed by quantum theory. One of the most convincing experiments was conducted in the 1980s by Alain Aspect (1947) and his colleagues at Orsay. It should be noted that the non-locality verified by these experiments does not contradict the principle of causality: EPR correlations cannot be used to transmit information instantaneously between two points (Haroche, 2002, p. 91).

on a crucial aspect of the problem), we will inevitably conclude that there is an entanglement between the atom and the cat system, which should be in a superposition of the live cat associated with the excited atom and the dead cat associated with the de-excited atom. Such a situation, which leaves the unfortunate cat suspended between life and death, was—as has already been stated—considered burlesque by Schrödinger[10].

Following Rae (1986), in the case of a photon passing through a polarizer horizontally, the cat will be unaffected. If the cat crosses the polarizer vertically, the camera will be activated, and the cat will die, but quantum physics states that only in the case where the box is opened and its state measured would the cat be alive and dead up to that moment. See the following figure:

[10] Haroche, 2002, p. 91.

**Figure 3**

*Schrödinger's Cat*

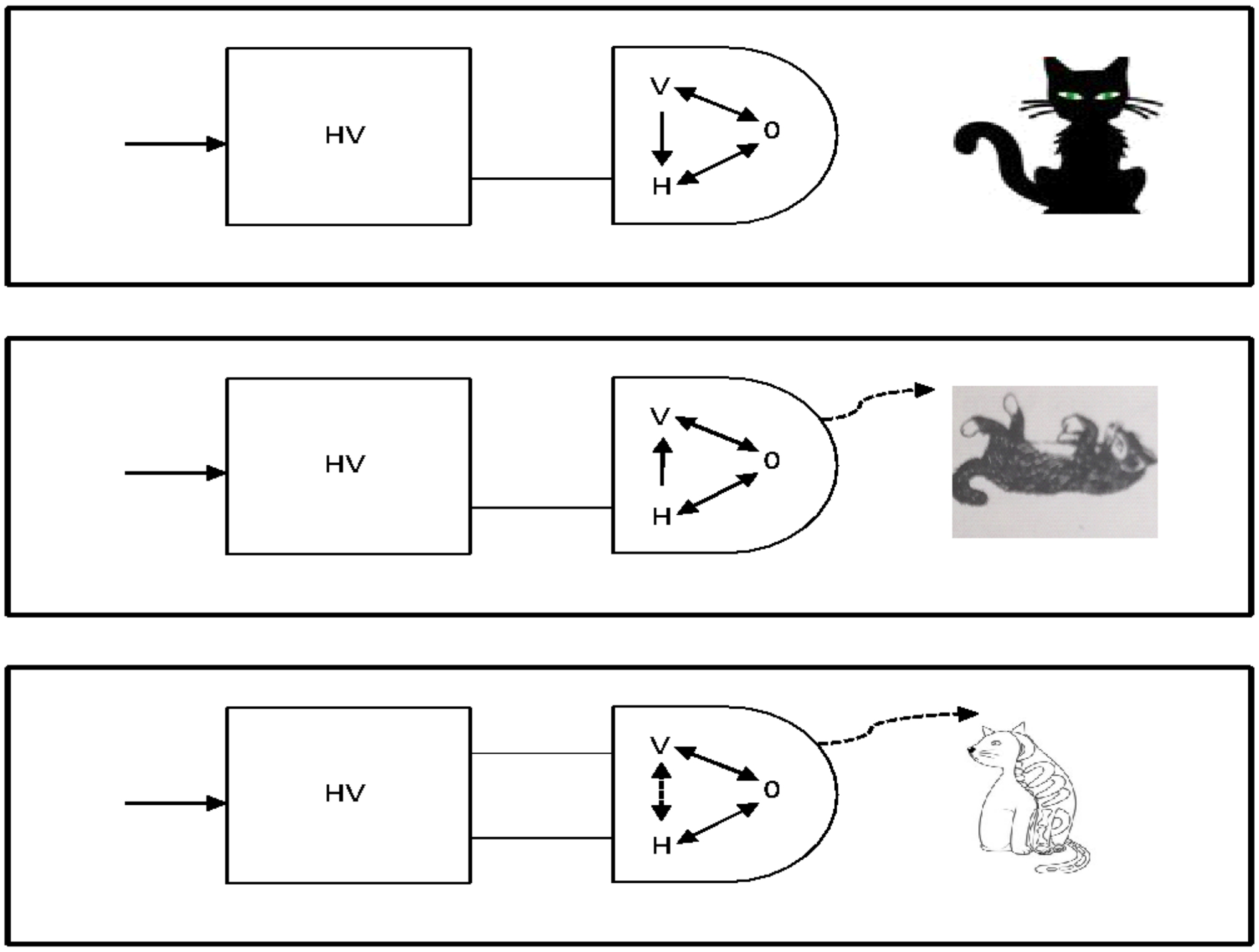

Source: Rae, 1986.

Much has been written about this problem. Some have said that it is now, when we try to observe whether the cat is alive or dead, that a *mental process* in the observer forces nature to *decide*. Others question whether the mental process of the cat itself must be considered, and the discussion then becomes more metaphysical. If we want to avoid this debate, Bohr's pragmatic approach is useful. To find out whether the superposition of states exists, we need to devise a specific observational device. The live-cat-dead cat superposition can only be demonstrated if we know how to design an experiment capable of revealing the interference of the complex amplitudes associated with the live and dead parts of the cat. Schrödinger did not take the discussion that far, but we could, for example, think of using a quantum mouse as a probe of the cat's state, and ask it to cross the box. The probability of the mouse escaping would then be the square of the sum of two amplitudes, one corresponding to the live cat and the other to the dead cat. Will we see an interference term in the final probability? This is unlikely and strongly contradicts our intuition (Haroche, 2002, p. 92).

The question is, what has happened to interference? Why has it disappeared? The answer has to do with the fundamental notion of decoherence. The situation we have schematized to the extreme has overlooked an essential element. Our cat cannot be isolated because it has neglected an essential element. Our cat cannot be isolated simply by considering a single atom deciding its fate. The cat—like any macroscopic system—is surrounded by an environment made up of numerous molecules and thermal photons, and its coupling with this environment cannot be neglected. If one wants to determine the path traveled by the particle, one needs, for example, to scatter a photon. This photon is then entangled with the particle, and entanglement destroys the coherence between the paths. If we measure the photon in one state, we know that the particle has gone through a hole. The other state is gone. There is no more interference. This allows us to better understand complementarity as an entanglement effect of the particle with itself. Our cat is in an analogous situation (Haroche, 2002, p. 93).

First, let us note that the starting point of our reasoning, the existence of a well-determined quantum state for the cat—leads to the existence of a well-determined quantum state of the environment. Even if we could decouple it from the rest of the world at this initial moment, it would be impossible to avoid its interaction with a bath of molecules and photons that would quickly find themselves in their different quantum states depending on whether the cat was alive or dead. Very quickly, information about the cat's state would leak into the environment, destroying the quantum interference, just as the photon scattered by the particle in Young's experiment makes the fringes disappear. The environment acts as a *spy*, eliminating the quantum ambiguity.

Finally, it should be noted that decoherence occurs increasingly rapidly as the size of the systems increases. This is because the larger the system, the more degrees of freedom it has that are coupled to the environment. It is not necessary to consider systems as macroscopic as a cat for decoherence to dominate. This is already the case for microscopic systems in the biological sense, such as macromolecules, viruses, or bacteria. Decoherence is just as effective in an inert object consisting of a large number of particles or an aggregate of atoms. The image of the cat is nothing more than an extreme metaphor imagined by Schrödinger to strike a chord (Haroche, 2002, p. 94).

## Quantum Decoherence Captured on Film

A clear example of decoherence can be seen in the figure below:

**Figure 4.**

*Experience of Decoherence*

a)

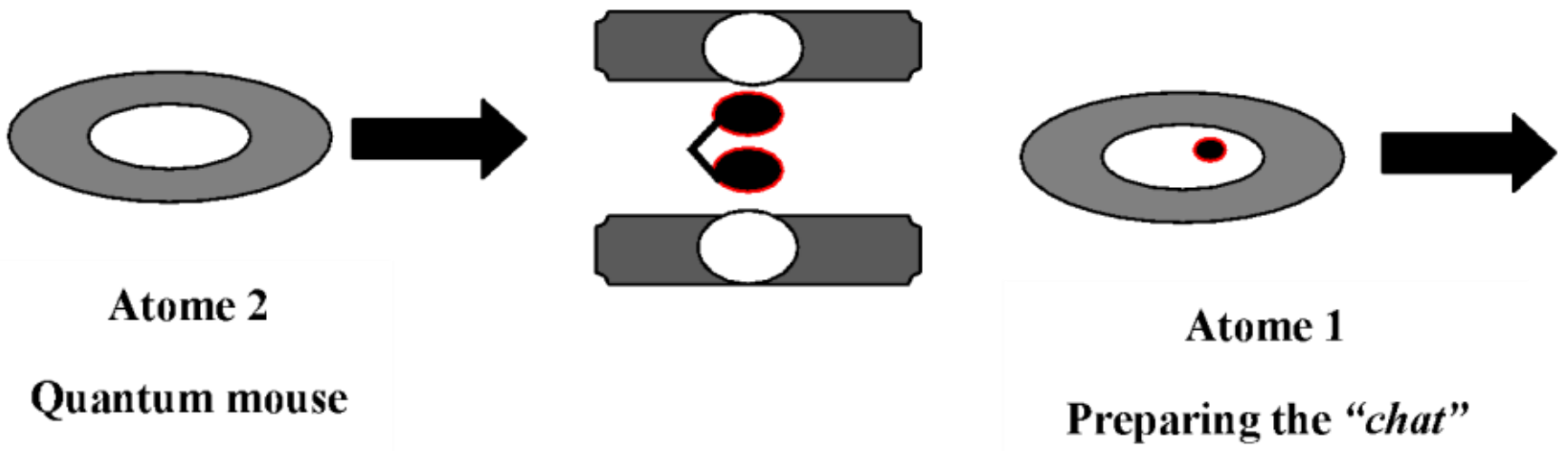

b)

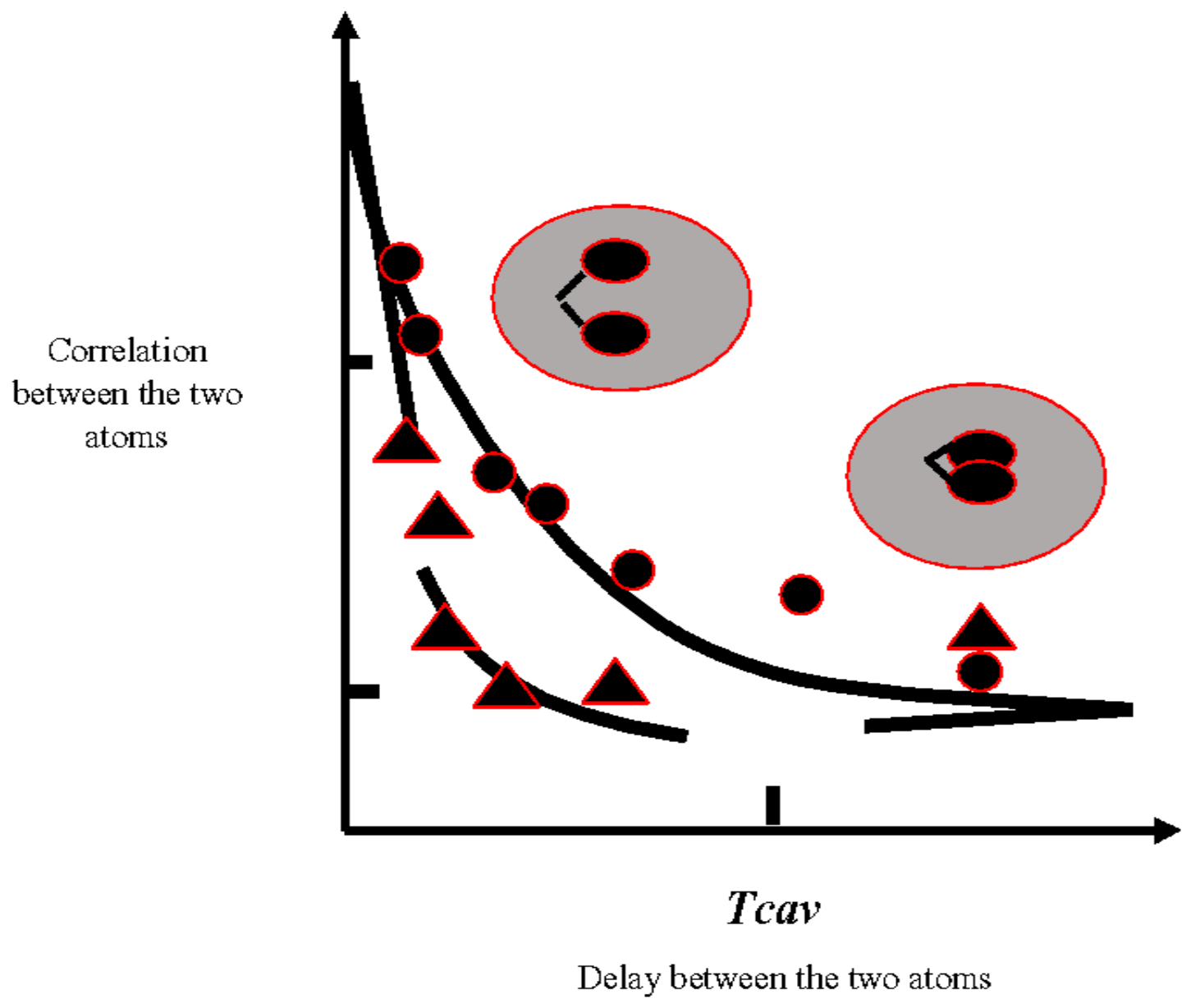

Source: Haroche (2002).

What can we say about the superposition of the states of the field itself? How long does this superposition of states last? The environment of the field is the space surrounding the cavity, which can be filled with photons scattered by the surface defects of the mirrors. It is this scattering process that limits the lifetime of the field in our experiment to a time $T_{cav\ time}$ of a fraction of a millisecond.

If the cavity contains *n* photons on average, a small field containing about one photon will escape into the environment in a brief time: $T_{cav\ time}$ divided by *n*. This microscopic field carries information about the phase of the field that remains in the cavity. Thus, after some time $T_{cav\ time}/n$, the quantum coherence between the two components of the field in the cavity has disappeared. This explains why macroscopic fields, for which n is exceptionally large (on the order of a million or more), behave classically, with decoherence being almost instantaneous.

The decoherence time is, therefore, long enough to allow transient observation of the quantum interferences associated with the two components of our *Schrödinger's cat*. For this observation, we send into the cavity, after the first atom that prepares the cat, a second atom that plays the role of the *quantum mouse* mentioned above. This atom recombines the field components separated by the first atom in such a way that, in a correlation signal between the results of the detections of the two atoms, an interference-sensitive term associated with the two components of the cat created by the first atom appears.

This interference signal decreases as the delay between the two atoms increases. The faster the two components of the *cat* are separated, the faster this phenomenon occurs, illustrating one of the essential aspects of decoherence, which acts all the more quickly the more *macroscopic* the system is. This experiment is an exploration of the boundary between the quantum world (in which interference effects are manifest) and the classical world (in which these effects are veiled).

The traditional view of Schrödinger's cat experiment considers that if we have a set of photons and a half-silvered mirror, which splits the impinging photon in a quantum state into a superposition of two different states, one of them is reflected, and the other passes through the mirror. There is a photon detected in the device in the path of the transmitted photon, which registers the arrival of a photon that is fired by a gun, which kills the cat.

The cat can be thought of as the endpoint of measurement. If one were to shift the perspective from the quantum level to the world of ponderable objects when the cat is found either alive or dead, there would be a direct operation of the observer. The problem, however, is that at the quantum level, to be logically consistent in reaching the cat level, one would have to believe that the current level of the cat's state is a superposition of both being dead and alive.

The main point is that the photon is in a superposition of states, going oner way or other. The detector is in a superposition of states—of being and not being—or, in the case of the cat, of being alive and dead (Penrose, 1997, p. 72). The mental exercise of the cat, or in this case, the observation of the behavior of photons, is an expression of the role of the observer in the description of microscopic phenomena.

Penrose (1997), following the words of Robert Wald (1947), who stated, *If you really believe in quantum mechanics, then you cannot take it seriously* (Penrose, 1997, p. 72), considers the division of quantum physics into several categories: those who believe and those who are serious, or those who consider that the state vector $|\psi\rangle$ describes the real world (i.e., the state vector represents reality) and those who do not consider it to be real. Those who believe in quantum mechanics, however, do not believe that this is a good attitude towards quantum mechanics. Penrose places several authors according to their position concerning these two categories.

For example, Niels Bohr and the Copenhagen School, in general, fall into the category of believers. Bohr was undoubtedly a believer in the ideas of quantum mechanics, but he did not take seriously the idea of the state vector as a valid description of reality (Penrose, 1997, p. 72). Somehow $|\psi\rangle$ was all in the mind—that is, it existed in our description of the world, even if it was not the world itself. According to Penrose, this is the same line that Bell follows in the so-called FAPP, as a reference to *For All Practical Purposes*. This diagram is divided into the so-called serious ones into other subcategories. Some believe that the U of the diagram is the complete history of quantum mechanics, a unique history with a distinguishable line of development and evolution. From this perspective, the idea of multiple worlds follows, in which view the cat is not only alive and dead but, in some sense, *inhabiting* different universes (Penrose, 1997, p. 74).

Those who consider the serious point of view —including Penrose himself—are those who believe that both U and R are real phenomena. There is not only a unitary evolution that takes place throughout the system. In this sense, even if it is something small, there is also something different that is directed outward, which is what Penrose calls R, which is not the R of the relation between the quantum level and the classical level, but something *very much like it* [11].

If one believes that one could take one of the two points of view, one could argue that there are no new effects in physics, and Penrose includes the Broglie/Bohm perspective. R plays a fundamental role in adding to the standard model of quantum mechanics U but without the expectation of finding new effects. However, some consider that the second point of view, the truly serious one, leads to a new change in the structure of quantum mechanics, where R contradicts U (Penrose, 1997, p. 72). The scheme presented by Penrose is as follows:

[11] It is uncertain whether it refers to decoherence *per se*.

**Figure 5**

*Perspective Believe vs Serious*

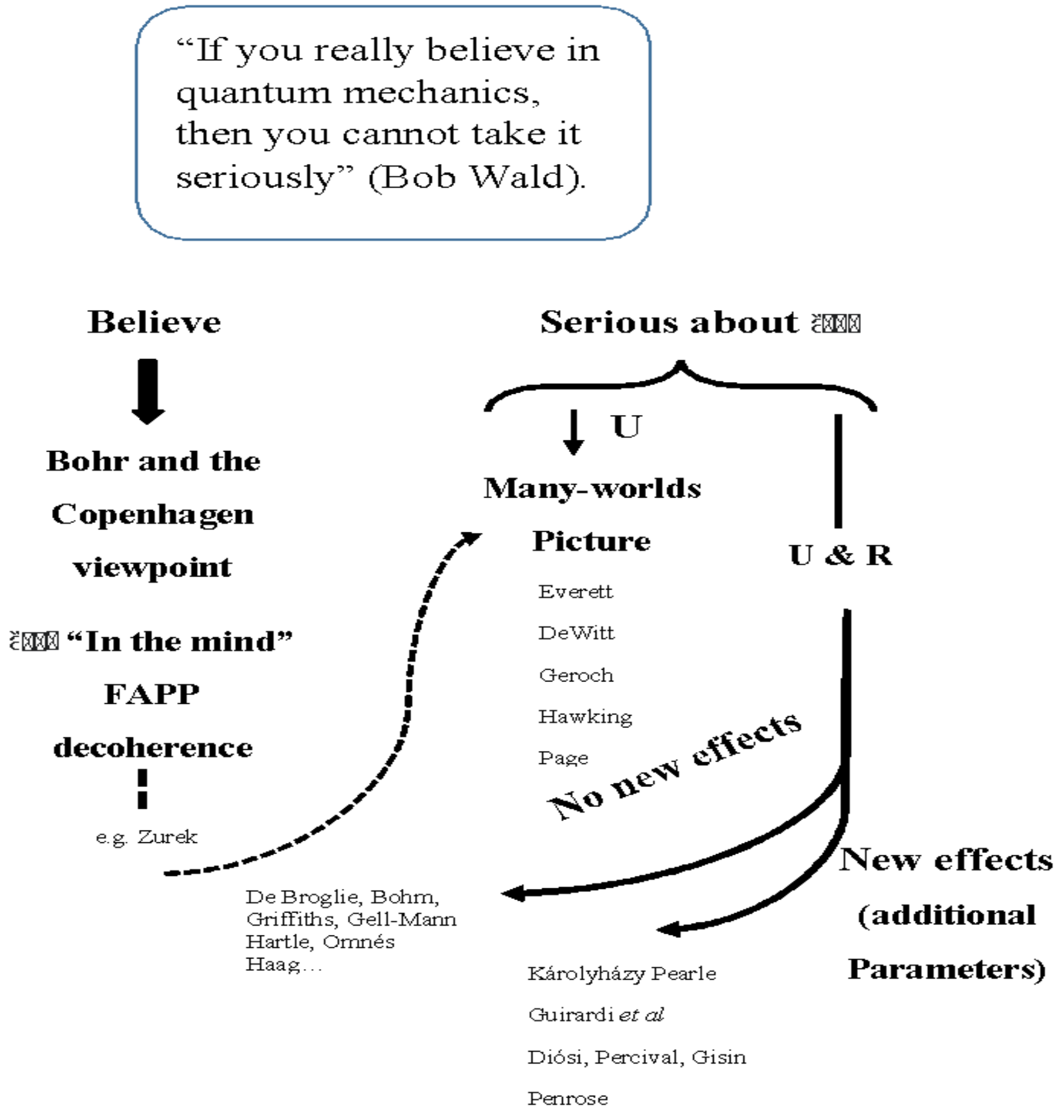

Source: Penrose (1997).

Penrose's idea is that the state vector perspective—if it offers a real explanation of the world and is not merely an abstract description rooted in a platonic-mathematical view disconnected from physical reality, which can be described, as will be seen later, as an element of existence related to consciousness or cognitive aspects— is different from those raised previously by Kuhn (2001) as internal elements or of adjustment and interpretation of reality.

## Summary

It is impossible to avoid the idea of analyzing the role of the observer from the point of view of physical phenomena without going through the analysis of the philosophical debate that focuses mainly on the different positions that support the existence of an external and independent world from the observer and those who focus on a perspective that considers that the world is a human construction, shaped by contextual elements and therefore directly affecting the observation of phenomena. In this sense, to be able to delimit the different perspectives that have been studied about *realism* and that we have tried to address briefly, they account for a latent discussion that will be observed much more deeply in several ideas put forward by quantum theories.

The role played by the observer in principles such as uncertainty or in what has come to be called the collapse of the wave function is precisely the result of this separation—which has philosophical roots—between physicists who argue for the fundamental role of the measuring apparatus and the observer and those who emphasize the fundamental role of the latter in *constructing* this interpretation of reality. Although the history of quantum mechanics itself shows how surprised the physicists who made these discoveries were— because the experiments contradicted classical mechanics—the debate is still open, and the impact on other ontological fields is evident.

The perspectives of articulation, such as those proposed by Penrose (1997), offer an important frame of reference for at least identifying the state of the discussion and have made it possible, as we have seen above, to introduce a discussion on the role of *consciousness* in the construction of this reality through the Popperian idea of the existence of the three worlds. The social sciences recognize the role of consciousness as fundamental in the consolidation of their fields of study and in the possibility for the social scientist to analyze, interpret, and construct reality.